\documentclass[sigconf]{acmart} 

\usepackage{tabularx}
\usepackage{multirow}
\usepackage{enumitem}
\usepackage{xcolor}
\usepackage{soul}
\usepackage[normalem]{ulem}

\newif\ifcomments

\commentstrue

\ifcomments

  \long\def\qj#1{%
    \begingroup
    \color{red}%
    [Georgie: #1]%
    \endgroup
  }

  \long\def\zl#1{%
    \begingroup
    \color{purple}%
    [Zhuoran: #1]%
    \endgroup
  }

  \long\def\yz#1{%
    \begingroup
    \color{blue}%
    [Yang: #1]%
    \endgroup
  }

  \long\def\yc#1{%
    \begingroup
    \color{orange}%
    [Yulin: #1]%
    \endgroup
  }

  \long\def\delete#1{%
    \begingroup
    \color{red}%
    \sout{#1}%
    \endgroup
  }

\else

  \long\def\qj#1{}
  \long\def\zl#1{}
  \long\def\yz#1{}
  \long\def\yc#1{}
  \long\def\delete#1{}

\fi

\AtBeginDocument{%
  }
\renewcommand\footnotetextcopyrightpermission[1]{}
\acmConference[]{}{}
\acmISBN{}

\begin{document}

\title{When Single-User-Oriented LLM-based Assistants Involve Others: A Scoping Review of Pathways, Risks, and Responses}

\author{Yulin Chen}
\orcid{0009-0009-8237-7197}
\affiliation{%
  \institution{Carnegie Mellon University}
  \city{Pittsburgh}
  \state{PA}
  \country{USA}
}
\email{yulinche@alumni.cmu.edu}

\author{Yang Zhan}
\affiliation{%
  \institution{North Carolina State University}
  \city{Raleigh}
  \state{NC}
  \country{USA}}
\email{yzhan3@ncsu.edu}

\author{Zhuoran Lu}
\affiliation{%
  \institution{University of Hong Kong}
  \city{Hong Kong SAR}
  \state{}
  \country{}}
\email{zhuoranl@hku.hk}

\author{Qiao Jin}
\affiliation{%
  \institution{North Carolina State University}
  \city{Raleigh}
  \state{NC}
  \country{USA}}
\email{qjin4@ncsu.edu}


\renewcommand{\shortauthors}{}

\begin{abstract}
  LLM-based assistants are increasingly extending into multi-party contexts, while core operational processes for context management, personalization, identity attribution, authority attribution, and action execution often remain organized around a single user. Existing work examines particular multi-party settings, but lacks a systematic account of how these single-user-oriented assistants begin to involve additional human parties and what risks emerge. To address this gap, we conducted a scoping review of 58 studies. We identify five operational pathways spanning direct and indirect involvement, five recurring risk domains, and five areas of implemented and proposed responses. Based on these findings, we argue for governance that attends to changing cross-person roles and relationships in practice, and for assistant designs that preserve person-specific boundaries throughout interaction.
\end{abstract}





\maketitle

\section{Introduction}

Powered by advances in large language models (LLMs), AI\footnote{We use ``AI'' broadly following OECD~\cite{organisation2022oecd} and NIST~\cite{ai2023artificial} formulations of AI systems as machine-based systems that generate outputs for given objectives and operate with varying levels of autonomy. Within this review, our scope is specifically LLM-based assistants whose interaction and behavior are substantially mediated by large language models.} assistants are becoming increasingly integrated into people’s everyday lives as long-term companions, personalized co-workers, and agents that carry out tasks on users’ behalf in the external world~\cite{wang_survey_2024, guan-etal-2025-survey, xi_rise_2025, zhang_survey_2025, qin_toolllm_2023}. Much of this work treats the individual user--AI assistant interaction as the basic unit of design. An assistant's tone and voice can be tailored to one user's preferences, recurring interactions can help it remember that user's habits and prior conversations, and more agentic assistants can plan or carry out tasks around that user's goals and requests~\cite{guan-etal-2025-survey,yang2026multi}. We refer to this organization of information and authority around a primary user as a \textit{single-user assumption}.

Despite this progress, the single-user assumption becomes increasingly difficult to sustain. Earlier research has already shown that intelligent assistants in settings such as families and organizations inherently encounter multiple users~\cite{dostreamerscare,smartspeakersreview,streamersandbystanders,youarebeingwatched}. More recent work shows that LLM-based assistants designed around an individual user are increasingly used in shared and collaborative settings, including family conversations and group planning among friends~\cite{shan2026pec,cheng2026grouptravelbench,zhang2025exploring}. Additional people can also become involved without directly using the assistant~\cite{zhan2026protecting,lee2026still}, as their speech or other information enters the assistant's context through shared environments or another user's resources. In these settings, information or permissions organized around one user can extend to others, creating consequential effects beyond the initiating user by exposing stored data to co-users, enabling sensitive inferences about non-users, or treating third-party speech as the primary user's own input~\cite{sen2026detecting,kim2026your,lee2026still}.

Existing research, however, captures only parts of this transition. Multi-party dialogue reviews focus mainly on computational challenges within conversations involving multiple speakers~\cite{ganesh-etal-2023-survey}; group conversational agent research examines systems intentionally designed to intervene in synchronous group interaction~\cite{10.1145/3800645.3812934}; and recent multi-user LLM work formalizes direct multi-principal settings involving conflicting objectives, authority, privacy, and coordination~\cite{yang2026multi}. These perspectives provide limited coverage of broader operational settings in which single-user-oriented assistant operation begins to involve other people through shared use, cross-user delegation, shared state or infrastructure, information about non-users, or sensing in shared environments. They also do not systematically connect these forms of involvement to their resulting risks, existing responses, and remaining gaps.

To address this gap, we conduct a scoping review of LLM-based assistants operating beyond the single-user assumption. We use \textit{multi-party operation} as an inclusive term covering both additional users who interact with an assistant and individuals who become involved indirectly through its information processing, sensing, resource use, or actions. We reviewed research published between 2022\footnote{The 2022 starting point aligns with prior reviews that situate the recent expansion of LLM-based conversational assistants around the public release and uptake of ChatGPT in late 2022~\cite{user_Autonomy,LLM-ification}.} and September 2026, resulting in a final corpus of 58 studies. We focus on cases where the assistant ordinarily organized around one user begins to involve additional human parties and produces an identifiable change in outcomes. This review makes three contributions to the HCI and AI communities. First, we distinguish the use contexts and actors involved in multi-party operation and identify five recurring operational pathways through which single-user-oriented assistants  reach additional human parties. Second, we synthesize five recurring domains of risk affecting primary users, additional users, and indirectly involved individuals, as well as relationships among them. Third, we map the identified risks to implemented and proposed responses, revealing uneven response coverage and unresolved gaps, and deriving implications for the design and governance of LLM assistants in multi-party contexts.

\section{Background}

\subsection{LLM-based Assistants and the Single-User Assumption}

Large language models have increasingly been incorporated into conversational assistants and agentic systems that extend beyond one-shot text generation. LLM-based assistants can maintain conversational context and persistent memory, reason over longer task sequences, invoke external tools and APIs, and perform actions on behalf of users~\cite{wang_survey_2024, qin_toolllm_2023, xu2025amemagenticmemoryllm}. These capabilities have expanded the role of LLMs from conversational interfaces toward assistants that support information seeking, planning, decision making, workplace tasks, and other forms of delegated activity~\cite{wang_survey_2024, yang2026multi}.

Despite this expansion in capability, many contemporary LLM assistants remain organized around an individual user. Conventional conversational systems have historically modeled interaction as a dyad between one user and one system~\cite{beyonddyadic, uxresearchliter, ganesh-etal-2023-survey}. Contemporary LLM interfaces largely inherit this structure. Standard chat templates typically represent human input through a single ``user'' role, while instruction tuning conditions the model on a unified conversational context and optimizes for an assistant response to that user~\cite{yang2026multi}. Preference-based alignment further commonly reduces feedback to an aggregated or scalar objective rather than explicitly representing multiple independently situated users with distinct objectives and authority~\cite{ouyang2022training, yang2026multi}. 
Our review builds on this account of the single-user assumption to examine how it is challenged as assistant operation extends beyond one user to involve additional human parties.

\subsection{Prior Reviews of AI Beyond Single-User Interaction}

Prior reviews have examined conversational AI beyond dyadic or single-user interaction from several perspectives. Ganesh et al.~\cite{ganesh-etal-2023-survey} focus on computational multi-party dialogue modeling, including speaker/addressee recognition, turn-taking, disentanglement, response generation, and information flow, but largely cover pre-LLM approaches. Yeo et al.~\cite{10.1145/3800645.3812934} review agents that sense and intervene in group interaction, with LLMs treated alongside rule-based and Wizard-of-Oz systems. Lei et al.~\cite{lei2026beyond} trace multi-party dialogue toward LLM-driven and multi-agent systems, but remain organized around MPD tasks, methods, datasets, and evaluation. Yang et al.~\cite{yang2026multi} more directly examine the transition from single- to multi-principal LLM agents, emphasizing instruction conflict, access control, and coordination in shared-agent settings.

These lines of work leave a broader transition less systematically examined. Multi-party dialogue research generally begins with a conversation that already contains multiple speakers~\cite{ganesh-etal-2023-survey, lei2026beyond}, while group conversational agent research~\cite{10.1145/3800645.3812934} begins with agents intentionally designed for group-level intervention. Recent multi-principal LLM work~\cite{yang2026multi} comes closer to our framing, but primarily considers multiple direct users interacting with a shared agent. In contrast, this review starts from LLM assistants organized around a single user and examines what happens when aspects of their operation begin to involve additional human parties. This perspective also captures indirectly involved individuals and connects these extensions to their resulting risks and corresponding technical or design responses.

\subsection{Conceptual Scope of This Review}

In this review, we use \textit{an LLM-based assistant} to refer to an interactive AI system in which a large language model plays a central role in interpreting user input and contextual information, generating responses or plans, or coordinating actions, tools, and resources on behalf of a user. This scope includes conversational assistants and more agentic forms of assistance that can maintain state, access external resources, or perform delegated actions. It does not refer to AI systems generally, or to algorithmic systems. Unless otherwise specified, the term \textit{assistant} throughout the remainder of this paper refers specifically to an LLM-based assistant in this sense.

We also distinguish \textit{multi-user}, \textit{non-user}, and \textit{principal}. We use \textit{multi-user} for settings in which multiple people directly use, access, address, or act through an assistant. A \textit{non-user} does not directly operate the assistant but becomes implicated through represented information, sensing in a shared environment, or exposure to assistant outputs or actions. Following prior work on multi-principal LLM agents~\cite{yang2026multi}, \textit{principal} is used to refer to a user whose objectives, instructions, authority, or resources the assistant is expected to represent or act upon.

Rather than treating multi-party AI as a predefined system category, this review starts from LLM assistants ordinarily organized around a single user and examines their transition into operation involving additional human parties. We use \textit{operation} broadly to include interaction and context processing, inference and sensing, resource or tool access, output generation, and delegated actions. Our central unit of analysis is this single-user-to-multi-party transition and its identifiable consequences for the primary user, additional direct users, indirectly involved individuals, and their relationships, allowing responses to be examined in relation to the specific cross-person operation and risks they address.

\section{Method}

We conducted a scoping review of how LLM-based assistants organized around a primary user extend beyond that user to involve additional human parties. We followed Arksey and O'Malley's five-stage scoping-review framework~\cite{Arksey01022005}, covering research-question development, study identification and selection, data charting, and synthesis. Study identification combined database searches with a supplementary Google Scholar search and backward and forward snowballing following Wohlin~\cite{wohlin2014guidelines}. We used the PRISMA Extension for Scoping Reviews (PRISMA-ScR)~\cite{tricco2018prisma} to guide transparent reporting of the search, study selection, data charting, and synthesis procedures.

\subsection{Research Questions}

The review was guided by the following research questions:

\begin{itemize}

    \item \textbf{(RQ1)} How do multi-party contexts extend AI assistant operation beyond the single user?
    
    \item \textbf{(RQ2)} What risks are reported when AI assistants operate in multi-party contexts beyond the single user?
    
    \item \textbf{(RQ3)} What design responses have been implemented or proposed to address these risks, and what research gaps remain?
\end{itemize}

\subsection{Databases and Corpus Selection}

\begin{figure*}[t]
    \centering
    \includegraphics[width=\textwidth]{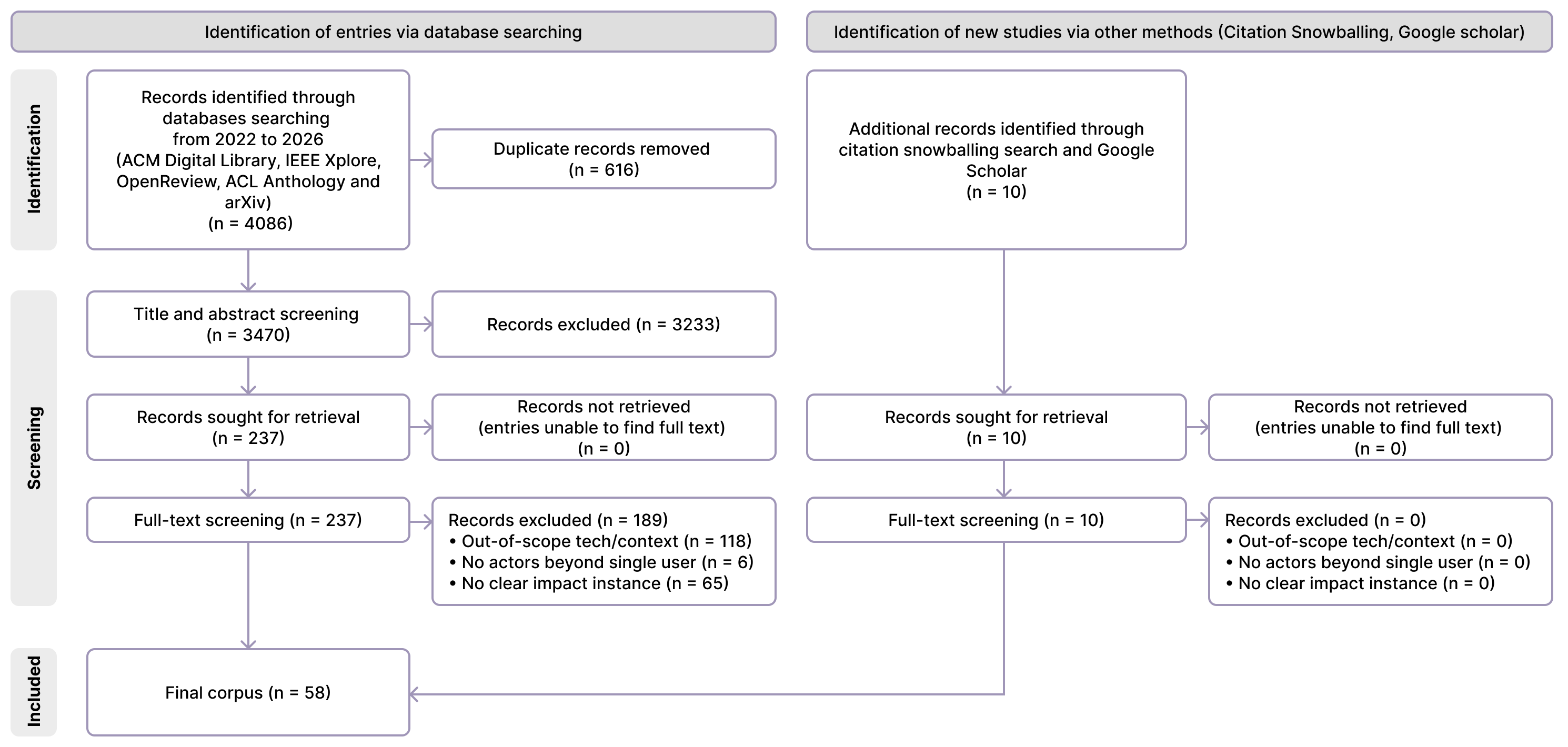}
    \caption{PRISMA-ScR flow diagram of study identification through database searching and supplementary methods.}
    \label{fig:sr_chart}
\end{figure*}

\subsubsection{Databases}

Research on LLM-based assistants in multi-party contexts spans HCI, NLP, AI and machine learning, security, robotics, and agent systems. We therefore searched five complementary sources: ACM Digital Library for HCI, CSCW, interaction design, and social computing; IEEE Xplore for AI systems and robotics; OpenReview for major machine-learning and related AI venues; ACL Anthology for broader archival coverage of NLP, conversational systems, and language-agent research not fully represented in OpenReview; and arXiv for rapidly developing preprint research across these areas.

We considered research from January 2022 to September 2026. The starting point was chosen to capture the emergence of publicly accessible LLM-based conversational assistants, particularly around the release of ChatGPT in late 2022, when such systems began to enter widespread use~\cite{user_Autonomy, LLM-ification}.

\subsubsection{Search Strategy}

The search strategy was organized around two conceptual sets corresponding to the core scope of the review: \textit{AI assistant systems} and \textit{actors and contexts beyond the single user}. 
Because terminology in this emerging area varies among HCI, NLP, and AI research, we expanded both sets to capture alternative technical terms and descriptions used in the literature.

\paragraph{AI assistant systems.}
We identified core terms including \textit{AI assistant}, \textit{large language model}, \textit{chatbot}, and \textit{conversational agent}, and expanded them to include related descriptions such as voice assistants, intelligent assistants, AI agents, generative AI, and GPT-based systems. Broader terms such as ``AI'' and ``artificial intelligence'' were retained because some studies describe assistant-like systems using general AI terminology.

\begin{quote}
\small
\textbf{Set A:} (chatbot OR ``conversational agent'' OR ``conversational agents'' OR
``intelligent assistant'' OR ``voice assistant'' OR ``AI system*'' OR
``AI assistant*'' OR ``AI agent*'' OR ``LM agent'' OR
``large language model'' OR LLM OR ``LLM-based'' OR GPT OR
``speech language model'' OR ``language model'' OR ``generative AI'' OR
AI OR ``artificial intelligence'' OR
``AI-powered'' OR ``AI-based'' OR ``AI-driven'' OR ``AI-assisted'' OR ``AI-enabled'')
\end{quote}

\paragraph{Actors and contexts beyond the single user.}
The second set captured situations where AI assistants involve actors beyond the single user, including direct participation such as multi-user and collaborative use, and indirect involvement such as bystanders, third parties, or individuals whose information enters shared interactions or contexts. Terms related to absent consent or unauthorized access were included to capture non-voluntary or unauthorized involvement that may not be described using explicit multi-user terminology. These terms identified cross-person involvement rather than pre-specifying risks or outcomes.

\begin{quote}
\small
\textbf{Set B:} (bystander* OR ``non-user*'' OR ``non primary user*'' OR ``non-primary user*'' OR
``third party'' OR ``third-party'' OR ``third parties'' OR
``multi-user'' OR multiuser OR ``multi user'' OR ``multiple users'' OR
``multi-party'' OR ``multiparty'' OR ``multiple speakers'' OR
``shared account*'' OR ``account sharing'' OR ``shared use'' OR ``shared access'' OR
``shared context'' OR ``shared conversation*'' OR ``shared history'' OR
``group interaction'' OR ``group decision'' OR
``collaborative use'' OR ``collaborative interaction'' OR
``without consent'' OR ``lack of consent'' OR ``unauthorized access'' OR
``secondary user*'' OR ``affected user*'' OR passersby OR
``second-hand'' OR ``non-consenting'' OR ``multi-tenant'')
\end{quote}

The two sets were combined using the Boolean expression:

\begin{quote}
\centering
\textbf{(Set A) AND (Set B)}
\end{quote}

Equivalent Boolean structures were adapted to the syntax and field restrictions supported by each source. Searches were primarily conducted over abstracts in ACM Digital Library, IEEE Xplore, ACL Anthology, OpenReview, and arXiv. The searches produced 4,086 records in total: 395 from ACM Digital Library, 1,456 from IEEE Xplore, 91 from ACL Anthology, 427 from OpenReview, and 1,717 from arXiv.

After consolidating the search results and removing duplicate records, 3,470 records proceeded to title and abstract screening.

\subsection{Screening Criteria}

Following deduplication, we screened the retrieved records according to the eligibility criteria and procedures described below.

\subsubsection{Title/Abstract Screening Criteria}

At the title and abstract stage, records were screened collaboratively by three authors using broad exclusion criteria intended to preserve recall. All records were assessed against the same predefined criteria, with ambiguous cases or differing interpretations discussed collectively until agreement was reached. The authors iteratively aligned their interpretation of the criteria throughout screening. Records that could not be clearly excluded from the title and abstract alone were retained for full-text assessment.

We applied the exclusion criteria in three steps. First, we excluded studies in which AI or language models served only as analytical tools rather than as the assistant system under study, such as using models to classify, predict, extract, or otherwise analyze data. Second, among studies that examined AI systems themselves, we excluded work concerned only with model training, inference, or foundational capability evaluation without a human-grounded assistant interaction. Third, among studies involving human--AI interaction, we excluded records that showed no reasonable indication that the assistant extended beyond a single user to involve additional human parties. Studies limited to interactions among AI agents or models without corresponding other relevant human parties were excluded on the same basis.

Systems, benchmarks, datasets, and evaluations explicitly designed for multi-user or multi-party assistant settings were retained for full-text assessment even where a specific multi-party risk was not apparent from the title or abstract, because the relevant evidence chain could only be assessed reliably from the full text.

Of the 3,470 records screened, 3,233 were excluded, leaving 237 for full-text assessment.

\subsubsection{Full-text Screening Criteria}

Full-text eligibility was assessed by the same three authors using the collaborative procedure described above and the three cumulative criteria below. Publications failing any criterion were excluded. For reporting exclusion counts, each excluded publication was assigned a single primary reason based on the first criterion it failed.

\begin{enumerate}
    \item \textbf{Single-User-Oriented LLM Assistant ($n=118$).}
    The study did not involve an LLM assistant or core assistant mechanism ordinarily organized around a single user. This could be reflected in a single owner of memory, resources, and authority; or a single personalization profile, preference set, and task objective.

    \item \textbf{Extension into a Human Multi-Party Operational Context ($n=6$).}
    The study did not show that at least one additional human party entered the assistant's operational context, causing an assistant ordinarily organized around one user to involve multiple people. This included direct multi-user interaction and non-user actors, whose information or behavior became part of the assistant's operation. 

    \item \textbf{Identifiable Evidence-Supported Multi-Party Risk ($n=65$).}
    The study did not provide sufficient evidence or reasoning to establish a traceable connection between an AI assistant’s operation involving multiple human parties and an observed, demonstrated, or anticipated consequence. Support could come from the publication’s own empirical findings, system evaluation, participant accounts, problem framing, design rationale, or conceptual argument. Studies that solely introduced mechanisms intended to adapt LLM assistants to multi-user or multi-party settings were excluded if they did not provide a complete risk instance connecting the single-user assumption, additional human involvement, cross-person operation, and an identifiable risk. The risk did not need to be the publication's primary focus, but unsupported statements or summaries of prior work were also insufficient.

\end{enumerate}

After full-text assessment, 189 of 237 publications were excluded. This resulted in 48 studies from the database search.

\subsubsection{Identification of new studies via other methods}

To broaden the search scope and reduce the likelihood of missing relevant literature, we complemented database searching with targeted Google Scholar searches~\cite{tricco2018prisma,10.1145/3800645.3812934, PRISMAupdate} and iterative backward and forward citation searching~\cite{wohlin2014guidelines}.

\textbf{Google Scholar search.} We conducted targeted Google Scholar searches as a supplementary identification method, which identified 3 eligible studies. 

\textbf{Citation snowballing.} We conducted iterative backward and forward citation searching on studies satisfying the full-text inclusion criteria. Backward citation searching examined their reference lists for potentially relevant earlier work, while forward citation searching identified subsequent publications that cited them. Newly identified records were evaluated using the same full-text screening criteria as the database-retrieved corpus, and eligible studies identified in each round were added as seeds for the next round. This iterative process continued for three rounds and stopped when no additional eligible studies were identified. Citation snowballing identified 7 eligible studies. 
Google Scholar searching and citation snowballing contributed 10 additional unique studies to the final corpus.

\subsubsection{Final corpus.} 
Combining studies identified through database searching and these supplementary methods resulted in a final corpus of 58 studies.

\begin{table*}[t]
\centering
\caption{Codebook analytical dimensions used to systematically analyze the included risk instances. Gray-shaded entries indicate predefined categories coded deductively, while unshaded entries were developed inductively through open coding.}
\label{tab:codebook}

\small
\setlength{\tabcolsep}{3pt}
\renewcommand{\arraystretch}{1.18}

\begin{tabularx}{\textwidth}{
>{\raggedright\arraybackslash}p{2.8cm}
>{\raggedright\arraybackslash}X
}
\hline

\textbf{Dimension} &
\textbf{Description} \\

\hline

\textbf{Context} &
Captures the social, physical, organizational, or task setting in which cross-person assistant operation occurs.
\begin{itemize}[leftmargin=*,nosep]
    \item Examples include households, vehicles, meetings, workplaces, collaborative spaces, enterprise environments, and public or shared physical environments.
\end{itemize}
\\

\hline

\textbf{Actor Description} &
Captures the human parties involved in the cross-person operation and their roles or relationships to the primary user and assistant.
\\

\colorbox{gray!20}{\textbf{Actor Type}} &
Captures the human parties involved in the cross-person operation, their relationship to the assistant, and whether their involvement is direct or indirect.
\begin{itemize}[leftmargin=*,nosep]
    \item \textit{Direct involvement}: additional actors intentionally use, address, contribute to, or act through the assistant.
    \item \textit{Indirect involvement}: additional actors become implicated through represented information, shared environments, or exposure to AI-mediated outputs or actions without directly operating the system.
\end{itemize}
\\

\hline

\textbf{Operational Pathway} &
Captures the recurring ways in which assistants designed around a single user extend to additional human parties, synthesizing how cross-person operation varies by context and actor involvement.
\\

\hline

\textbf{Risk Description} &
Captures the risk that arises when an assistant originally organized around a single user involves additional human parties.
\\

\colorbox{gray!20}{\textbf{Risk Evidence Type}} &
Captures whether the risk was supported by the study's own empirical evidence or asserted conceptually by the authors.
\begin{itemize}[leftmargin=*,nosep]
    \item \textit{Empirical}: supported by evidence generated within the study, including user studies, interviews, benchmarks, experiments, or system evaluations.
    \item \textit{Conceptual}: articulated by the authors through the study's motivation, problem framing, discussion, or conceptual analysis without direct empirical support from the study itself.
\end{itemize}
\\

\hline

\textbf{Response Description} &
Records the specific intervention associated with the risk instance and the point at which it acts on the cross-person operation.
\\

\colorbox{gray!20}{\textbf{Response Type}} &
Captures the status of the response associated with a risk instance.
\begin{itemize}[leftmargin=*,nosep]
    \item \textit{Implemented response}: a concrete mechanism, architecture, training method, interface feature, or safeguard was implemented in the study.
    \item \textit{Future direction only}: a design recommendation, conceptual mitigation, architectural direction, or future research direction was proposed without a corresponding implemented response.
    \item \textit{No recorded response}: no implemented or proposed response was identified for the corresponding risk instance.
\end{itemize}
\\

\hline

\end{tabularx}
\end{table*}

\subsection{Data Analysis}

We conducted a deductively structured thematic analysis~\cite{Braun01012006} with inductive refinement of lower-level categories. Deductive qualitative analysis began with an explicit analytical scheme while allowing categories to be refined as the scheme is applied to the data~\cite{robinson2026deductive,pearse2019illustration}. This approach suited our review because we began with a specific conceptual transition of interest: how the LLM-based assistant organized around a single user operates once additional human parties become involved.

We used a \textit{risk instance} as the primary unit of analysis. A risk instance was defined as the smallest traceable case in which an assistant, ordinarily organized around a single user, came to involve at least one additional human party, and this cross-person operation was associated with an identifiable risk. Each instance therefore contained a traceable relationship among the single-user assumption, additional human involvement, the relevant cross-person operation, and the resulting risk, and additionally recorded whether a corresponding response was implemented, proposed, or not identified. Within a study, cases were coded as separate instances when they represented different risk relationships, such as a distinct consequence or a materially different form of cross-person operation leading to risk. Multiple examples, participants, experimental conditions, and passages describing the same underlying risk relationship were consolidated into a single instance. Differences in context, actors, pathways, or responses alone did not automatically constitute separate instances.

Based on our research questions, we structured the codebook around five substantive areas: \textit{context}, \textit{actors}, \textit{operational pathway}, \textit{risk}, and \textit{response}. Within this structure, three categorical dimensions were predefined and coded deductively using closed coding. \textit{Actor Type} distinguished direct from indirect involvement, informed by the NIST AI Risk Management Framework~\cite{ai2023artificial}, the OECD framework~\cite{organisation2022oecd}, and our conceptual scope. \textit{Risk Evidence Type} distinguished risks supported by the study's own empirical evidence from those articulated conceptually by the authors. \textit{Response Type} classified responses as implemented, proposed as a future direction, or not recorded.

The remaining dimensions were developed inductively through open coding. \textit{Context} captured the setting in which cross-person operation occurred, while \textit{Actors Description} recorded the human parties involved and their roles or relationships to the primary user and assistant. \textit{Operational Pathway} codes were developed by comparing recurring ways in which assistant operation extended to additional human parties under different contextual and actor configurations. \textit{Risk Description} and \textit{Response Description} retained the specific consequence and corresponding intervention identified in each instance. As coding progressed, these open codes were compared, consolidated, and refined into recurring categories used in the RQ1--RQ3 synthesis. The final codebook therefore comprised eight analytical dimensions shown in Table~\ref{tab:codebook}; gray-shaded dimensions indicate predefined closed-code categories, while the remaining dimensions were developed through open coding.

Coding followed a staged collaborative review process. One researcher conducted the initial extraction and coding of risk instances using the evolving codebook. The other two researchers then reviewed the extracted instances and coding decisions against the supporting textual evidence. Cases involving uncertainty about risk-instance inclusion, instance boundaries, or category assignment were discussed by all three researchers and revised until agreement was reached. The final codebook and coded dataset were established through this reconciliation process.

\section{Corpus Overview}

\subsection{Year of Publication}

The final corpus comprised 58 papers published between 2023 and 2026. Two studies were published in 2023~\cite{tan2023chatgpt,addlesee2023multi}, and three in 2024~\cite{zhan2024beyond,cheng2024cibenchbenchmarkingcontextualintegrity,shao2024privacylens}. Fifteen studies were published in 2025~\cite{jhamtani2025llm,mccarthy2026ethics,rezazadeh2025collaborative,lee2025map,bhatt2025enterprise,patlan2025murmur,song2025beyond,inoue2025llm,garcia2025evaluating,liu2025proactive,zhang2025exploring,wang2025multi,ghalebikesabi2025privacy,lan2025contextual,mireshghallah2026cimemories}. The remaining 38 studies were published in 2026~\cite{srivastava2026listening,jin2026security,malecot2026harmoni,yang2026no,yang2026multi,yang2026clawnet,al2026afa,cheng2026grouptravelbench,fan2026harness,ruzzetti2026muppet,wang2026weclawarena,hu2026evaluating,wang2026voxsafebench,shan2026pec,zhan2026protecting,sen2026detecting,kim2026your,song2026don,chen2026prism,ren2026gatemem,chen_if_2026,xu2026ai,lutz2026edge,gupta2026pisas,wang2026voxprivacy,li2026privacy,zhu2026choose,wang2026agentsocialbench,lee2026still,yang2026groupmembench,mitra2026adaptive,li2026whose,chen2026vehiclemembench,shapira2026agentschaos,bhagtani2026speakstaysilentcontextaware,vijayvargiya2026openagentsafety,wei2026clawsafetysafellmsunsafe,goel2026capablecarelesscomputeruseagents}. Publication activity increased sharply after 2024, with 53 of 58 studies (91.4\%) published in 2025--2026 and 65.5\% in 2026 alone, despite the September 2026 cutoff.

\subsection{Publication Venues}

Of the 58 included studies, 30 (51.7\%) were listed as arXiv preprints. The remaining 28 (48.3\%) were archival publications, comprising 23 conference papers, two workshop papers, and three journal articles. Among the archival venues, CHI was the most represented ($n=4$). Findings of ACL, Findings of EMNLP, CHI Extended Abstracts, and ICLR each contributed two studies. The remaining archival studies were distributed across HCI, NLP, AI/ML, security, VR, and agent-system venues.

\subsection{Risk Instances Overview}

We identified 118 distinct risk instances from the 58 included studies, with an average of 2.03 instances per study ($median=2$). The number of instances contributed by an individual study ranged from one to eight. One study contributed the maximum of eight instances~\cite{shapira2026agentschaos}.

The evidential basis of these risk instances also varied. Ninety-four instances (79.7\%) were supported by empirical evidence from the study itself, including user studies, benchmarks, experiments, and system evaluations. The remaining 24 instances (20.3\%) were conceptual risks, identified through motivating examples, conceptual analysis, discussion, or author interpretation rather than directly demonstrated through the study's own empirical results.

\begin{table*}[t]
\centering
\caption{Operational pathways through which LLM assistants extend beyond a single user. Pathways were coded at the risk-instance level. A study may contribute multiple risk instances assigned to different pathways.}
\label{tab:multi_party_contexts}

\small
\setlength{\tabcolsep}{2.5pt}
\renewcommand{\arraystretch}{1.30}

\begin{tabularx}{\textwidth}{
p{1.55cm}
p{2.65cm}
p{2.35cm}
>{\raggedright\arraybackslash}X
>{\raggedright\arraybackslash}X
}
\hline

\textbf{Involvement Mode} &
\textbf{Operational Pathway} &
\textbf{Actors} &
\textbf{Description} &
\textbf{Includes} \\
\hline

\multirow{3}{1.55cm}{\textbf{Direct AI Involvement}}
&
\textbf{Shared and Collaborative AI Assistant Use} &
AI co-users &
Multiple people directly use, access, or interact with the same AI assistant, including through shared conversations or accounts. &
\cite{jin2026security,rezazadeh2025collaborative,malecot2026harmoni,yang2026no,yang2026multi,al2026afa,cheng2026grouptravelbench,fan2026harness,ruzzetti2026muppet,hu2026evaluating,wang2026voxsafebench,shan2026pec,lee2025map,song2026don,chen2026prism,patlan2025murmur,ren2026gatemem,xu2026ai,song2025beyond,wang2026voxprivacy,tan2023chatgpt,addlesee2023multi,inoue2025llm,garcia2025evaluating,li2026privacy,liu2025proactive,zhang2025exploring,yang2026groupmembench,mitra2026adaptive,chen2026vehiclemembench,wang2025multi,bhagtani2026speakstaysilentcontextaware}
\\

\cline{2-5}

&
\textbf{AI-mediated Cross-user Interaction} &
AI agent user, other users or their agents &
A user-associated assistant communicates, coordinates, represents, or acts beyond its associated user's context with other people or their agents. &
\cite{jhamtani2025llm,jin2026security,yang2026clawnet,wang2026weclawarena,gupta2026pisas,chen_if_2026,zhu2026choose,wang2026agentsocialbench,li2026whose,ghalebikesabi2025privacy,shapira2026agentschaos,vijayvargiya2026openagentsafety,cheng2024cibenchbenchmarkingcontextualintegrity,wei2026clawsafetysafellmsunsafe,lan2025contextual,mireshghallah2026cimemories,shao2024privacylens,goel2026capablecarelesscomputeruseagents}
\\

\cline{2-5}

&
\textbf{Shared AI Infrastructure} &
Users with distinct accounts, permissions, or data scopes &
Separate users become operationally connected through shared AI models, retrieval-augmented generation (RAG) systems, memory, resources, or other backend infrastructure. &
\cite{sen2026detecting,gupta2026pisas,bhatt2025enterprise}
\\

\hline

\multirow{2}{1.55cm}{\textbf{Indirect AI Involvement}}
&
\textbf{Information-mediated Involvement} &
Primary users, non-user data subjects &
Information about non-users enters the assistant's context through user-connected data and resources. &
\cite{zhan2024beyond,kim2026your,li2026whose,shapira2026agentschaos,shao2024privacylens,goel2026capablecarelesscomputeruseagents}
\\

\cline{2-5}

&
\textbf{Sensing-mediated Involvement} &
Primary users, nearby non-users &
Non-users enter the assistant's perceptual \& output context through sensing in shared physical or social environments. &
\cite{srivastava2026listening,mccarthy2026ethics,wang2026voxsafebench,zhan2026protecting,zhan2024beyond,chen2026prism,chen_if_2026,lutz2026edge,lee2026still}
\\

\hline
\end{tabularx}
\end{table*}

\section{RQ1: How Do Multi-Party Contexts Extend AI Assistant Operation Beyond the Single User?}
\label{sec:rq1}

To answer RQ1, we first provide a descriptive overview of the contexts in which cross-person assistant operation occurred and the roles through which additional human parties became involved. We then present the five operational pathways identified across these contexts, organized into direct and indirect forms of involvement. 

\subsection{Context and Actor Involvement Overview}

The 118 risk instances identified across the 58 included studies spanned a range of application and social contexts. They were most frequently situated in enterprise and organizational contexts ($n=31$), personal digital assistance and external communication ($n=25$), domain-unspecified group dialogue and collaboration ($n=21$), and shared physical and social environments ($n=18$). Other contexts included shared accounts, conversations, and customized assistants ($n=8$); healthcare and care ($n=7$); households and families ($n=6$); travel planning and booking ($n=5$); negotiation and commercial transactions ($n=5$); education ($n=2$); in-vehicle assistance ($n=1$); and shared cloud deployment ($n=1$). Context counts were non-exclusive because a risk instance could span more than one context.

Because some risk instances contained more than one actor, the actors and related involvement were summarized across 123 assignments. Direct involvement accounted for 100 assignments (81.3\%). Actors occupied several recurring roles in relation to the AI assistant. They included co-users and collaborators who directly interacted with an assistant ($n=56$); recipients and interlocutors who communicated with a primary user through assistant-mediated content or actions ($n=41$); members of households, organizations, or other groups sharing an assistant or its supporting infrastructure ($n=3$). 
Indirect involvement accounted for 23 assignments (18.7\%). These included individuals whose personal or relational information was represented in assistant operation under Information-mediated Involvement ($n=8$). The remaining cases fell under Sensing-mediated Involvement ($n=15$), including people captured through environmental sensing and those exposed to assistant-generated outputs or actions without directly operating the assistant.

\subsection{Direct-Involvement Pathways}
\label{sec:direct_involvement}

\subsubsection{Shared and Collaborative AI Assistant Use}
\label{sec:rq1 Shared and Collaborative AI Assistant Use}

Shared or co-used assistants arise when multiple people directly use, access, or interact with the same AI assistant, either simultaneously or at different times. Family members, friends, coworkers, or even strangers sharing a subscription could access the same account and conversation history~\cite{song2026don,zhang2025exploring}. Platforms also allowed one user to share a conversation or customized assistant that another person could access, continue, or reuse~\cite{li2026privacy}. Long-lived personal agents could similarly be accessed by several people through repeated sessions and connected messages, files, and other resources~\cite{jin2026security}.

Shared use also occurred when multiple people relied on one assistant for a common service, discussion, or task. In homes, care settings, and vehicles, a shared assistant served different users with distinct preferences, histories, and goals~\cite{shan2026pec,al2026afa,wang2026voxprivacy,malecot2026harmoni,chen2026vehiclemembench}. For example, household members could use the same smart-home LLM assistant to control devices according to different temperature preferences~\cite{shan2026pec,al2026afa,wang2026voxprivacy,malecot2026harmoni,chen2026vehiclemembench}. In conversational and collaborative settings, groups used a common assistant to seek information, participate in discussion, coordinate goals, and complete joint activities~\cite{addlesee2023multi,tan2023chatgpt,inoue2025llm,song2025beyond,wang2025multi,wang2026voxsafebench,chen2026prism,xu2026ai}. This included group planning and personalization systems that combined users' preferences and constraints~\cite{cheng2026grouptravelbench,lee2025map}, agents that received instructions or objectives from multiple users~\cite{yang2026multi,fan2026harness}, and shared-state or memory systems that retained messages, decisions, preferences, and contextual information contributed over time~\cite{rezazadeh2025collaborative,yang2026no,patlan2025murmur,ren2026gatemem,hu2026evaluating,yang2026groupmembench}.

\subsubsection{AI-mediated Cross-user Interaction}
\label{sec:rq1 AI-mediated Cross-user Interaction}

Unlike the shared use, this pathway preserves the association between an assistant and a particular user while extending its operation to other people.
It included information gathering, communication, negotiation, and other interactions conducted on the user's behalf. An assistant could contact colleagues to obtain information held by different people, as in workplace tasks where it identified and communicated with relevant collaborators on behalf of the initiating user~\cite{jhamtani2025llm}. In delegated activities such as negotiation, the assistant could communicate directly with another person while continuing to represent its principal~\cite{zhu2026choose,li2026whose}. Personal assistants could similarly interact with friends, family members, coworkers, customers, and other human recipients in social, organizational, and task-specific communication settings~\cite{chen_if_2026,jin2026security,shapira2026agentschaos,vijayvargiya2026openagentsafety,wei2026clawsafetysafellmsunsafe,ghalebikesabi2025privacy,cheng2024cibenchbenchmarkingcontextualintegrity,lan2025contextual,mireshghallah2026cimemories,shao2024privacylens,goel2026capablecarelesscomputeruseagents}.

The same pathway also appeared in interactions among user-associated agents. Cross-user agent networks connect assistants representing different principals, allowing information, requests, and approvals to move between otherwise separate user contexts while coordinating delegated actions across them~\cite{yang2026clawnet,wang2026weclawarena,gupta2026pisas,wang2026agentsocialbench}. Here, users did not share the same assistant. The multi-party connection arose through interaction among their respective assistants.

\subsubsection{Shared AI Infrastructure}
\label{sec:rq1 Shared AI Infrastructure}

Shared AI infrastructure connects users who otherwise retain separate accounts, permissions, data scopes, or interactions through common backend components. Unlike shared or co-used assistants, these users do not necessarily interact with the same conversational interface; unlike AI-mediated cross-user interaction, their assistants do not directly mediate communication between them. The multi-party connection instead arises because otherwise separate user contexts depend on common model services, RAG pipelines, APIs, databases, caches, or memory layers~\cite{sen2026detecting,bhatt2025enterprise,gupta2026pisas}.
%
For example, an LLM-powered application designed for single-user local use could be deployed on a shared cloud server, where separate user sessions rely on common backend resources. In one case, two users ran the same application in separate sessions while sharing a Redis cache that did not distinguish between them~\cite{sen2026detecting}. Other configurations involved organizational users relying on the same fine-tuned model or RAG pipeline~\cite{bhatt2025enterprise}, or separate tasks in multi-user agent systems accessing a common memory layer across user contexts~\cite{gupta2026pisas}.

\subsection{Indirect-Involvement Pathways}
\label{sec:indirect_involvement}

\subsubsection{Information-mediated Involvement}
\label{sec:rq1 Information-mediated Involvement}

Information-mediated involvement occurs when a person who does not directly use the assistant becomes represented through information available within another user's context. User-provided emails, messages, documents, images, and other materials can contain jointly held communications, identifiers, or personal information about friends, colleagues, family members, and other people~\cite{zhan2024beyond}. Delegated interactions can similarly include information about a third person within the principal's briefing or conversational context~\cite{li2026whose}.

This pathway also extends through the broader information sources available to personal agents. Access to messages, email, calendars, documents, and other connected resources can bring together distributed traces concerning people in the user's social network~\cite{kim2026your}. Mailboxes and personal histories can contain information about both the account owner and other people represented in their communications~\cite{shapira2026agentschaos,shao2024privacylens}, while computer-use agents can encounter third-party information in the user's applications and workspace~\cite{goel2026capablecarelesscomputeruseagents}. In these cases, the additional person enters the assistant's operation as an information subject rather than as a direct participant.

\subsubsection{Sensing-mediated Involvement}
\label{sec:rq1 Sensing-mediated Involvement}

Sensing-mediated involvement occurs when people who are not directly using the assistant enter its operational context through speech, visual information, behavior, or physical presence. This pathway appears when an assistant's perceptual context extends beyond the primary user.
Audio sensing can bring nearby speakers and ongoing conversations into assistant operation. AI note-taking assistants capture conversations among meeting participants~\cite{mccarthy2026ethics}, while proactive, wearable, and audio-enabled assistants process speech from interlocutors, bystanders, family members, or other nearby speakers~\cite{srivastava2026listening,zhan2024beyond,zhan2026protecting,lutz2026edge}. Voice assistants may also incorporate speaker characteristics, background voices, and acoustic context~\cite{wang2026voxsafebench}, including third-party interjections during an ongoing interaction~\cite{lee2026still}. 
Visual sensing similarly brings surrounding people and activities into multimodal assistant operation through images or video~\cite{zhan2024beyond}. In XR collaboration, headset cameras can expose collaborators, nearby people, objects, and the physical environment to multimodal assistants~\cite{chen2026prism}. Spatially situated embodied assistants likewise operate around co-present friends, colleagues, or strangers, extending their perceptual and interaction context beyond the primary user~\cite{chen_if_2026}.

\section{RQ2: What Risks Are Reported When AI Assistants Operate in Multi-Party Contexts Beyond the Single User?}
\label{sec:rq2}

We identified five risk domains based on where the primary harm or breakdown occurs for people. A single operational pathway could give rise to multiple risk domains. For each domain, we first describe the risks it contains and then briefly explain how these risks are associated with the operational pathways.

\subsection{Conversational Attribution and Turn-Taking Errors}
\label{sec:Conversational Attribution and Turn-Taking Errors}

Multi-party conversations produced errors in both participant attribution and conversational turn-taking. Once several people shared an interaction, assistants misidentified speakers and addressees, repeated answers already supplied by another participant, responded to human--human speech, or failed to determine whether they should take the conversational floor~\cite{tan2023chatgpt,addlesee2023multi,inoue2025llm,song2025beyond}. Reactive and pause-based turn-taking policies similarly became unreliable because silence, overlap, backchannels, and turn-yielding no longer uniquely indicated that the assistant should speak. This produced unnecessary prompting, unwanted entries, interruptions, and disruptions of human--human conversation~\cite{liu2025proactive,garcia2025evaluating,mitra2026adaptive,bhagtani2026speakstaysilentcontextaware}.

Attribution failures also affected whether participant information and interaction state remained associated with the correct person. Shared assistant context could associate one person's conversational history with another person, producing incorrect personalization in households, care settings, and shared vehicles~\cite{malecot2026harmoni,al2026afa,shan2026pec,chen2026vehiclemembench}. Assistants similarly lost distinctions among an individual speaker, a group of speakers, the participant to whom a reply should be directed, and audience-only listeners, causing information to be attributed to the wrong participant or the beliefs and terminology of multiple participants to be merged~\cite{hu2026evaluating,yang2026groupmembench}. Pairwise dialogue training also produced role and stance drift in longer multi-person exchanges~\cite{wang2025multi}. In spoken interaction, nearby or acoustically similar speakers could be mistaken for the primary user, allowing another person's utterance to alter an ongoing request or gain access to information associated with the intended speaker~\cite{lee2026still,wang2026voxprivacy}. 

As shown in Figure~\ref{fig:risk_connect_chart}, the shared or co-used assistant pathway (Section~\ref{sec:rq1 Shared and Collaborative AI Assistant Use}) is the main operational pathway associated with this risk domain, bringing multiple speakers into the same assistant interaction. Sensing-mediated involvement (Section~\ref{sec:rq1 Sensing-mediated Involvement}) also contributes by bringing nearby speakers and other people into interactions centered on a primary user.

\subsection{Cross-Person Information Boundary Violations}
\label{sec:Cross-Person Information Boundary Violations}

Cross-user access and persistence occurred when information associated with one user became accessible to another through shared accounts or common AI infrastructure. Shared accounts could expose prompts, generated responses, and other user traces to co-users~\cite{song2026don,li2026privacy}. In multi-party conversations, one inappropriate disclosure could expose private information to several participants at once, with models showing greater leakage than in comparable one-to-one settings~\cite{ruzzetti2026muppet}. Shared voice assistants could similarly reveal one user's sensitive disclosure when another user subsequently queried the same conversational context~\cite{wang2026voxprivacy}. Persistent multi-user memory allowed information contributed by one person to remain available during another person's later interaction, while deletion, redaction, and access-control mechanisms did not always remove or restrict information consistently for different users~\cite{rezazadeh2025collaborative,ren2026gatemem,gupta2026pisas}. Shared caches, fine-tuned models, and retrieval systems similarly connect otherwise separate user or organizational data scopes, making one user's or department's information available to another~\cite{sen2026detecting,bhatt2025enterprise}.

Assistants also violated information boundaries through disclosure to new recipients  because information available within one user--assistant relationship could become inappropriate in another. Personal agents that accessed email, documents, calendars, or other user resources could carry information from the primary user's private context into tasks involving other people~\cite{jhamtani2025llm,wang2026weclawarena,yang2026clawnet,wang2026agentsocialbench,li2026whose}. Information-sharing assistants similarly disclosed more personal information than was necessary or appropriate for an external recipient and task~\cite{ghalebikesabi2025privacy}. Contextual-integrity evaluations found that assistants made disclosure decisions inconsistent with context-specific informational norms, including both inappropriate disclosure and unjustified withholding of information that could legitimately be shared~\cite{cheng2024cibenchbenchmarkingcontextualintegrity,lan2025contextual,mireshghallah2026cimemories,shao2024privacylens,goel2026capablecarelesscomputeruseagents}. Persistent personalization further enlarged this exposure surface by retaining more user attributes that could later be reused in an inappropriate recipient context~\cite{mireshghallah2026cimemories}.

Information boundary violations also arose through inference and sensing involving non-users. Personal archives allowed agents to combine distributed traces from messages, emails, calendars, and documents to make sensitive inferences about related people~\cite{kim2026your}. User-provided conversations, images, video, and other resources could similarly contain information about non-user subjects that became available to the assistant through another person's context~\cite{zhan2024beyond,shao2024privacylens}. Sensing created a parallel route: audio and visual capture components of multi-modal assistants captured nearby speech, facial imagery, activities, or other sensitive information and processed or transmitted them as part of assistance for the primary user~\cite{srivastava2026listening,mccarthy2026ethics,zhan2026protecting,chen2026prism,lutz2026edge,wang2026voxsafebench}. In these cases, information originally embedded within another person's activities, relationships, or physical surroundings became inferable, persistent, or actionable within someone else's AI interaction.

All five operational pathways identified in RQ1 could give rise to risks in this domain by allowing information associated with one person to become accessible, persistent, inferable, or actionable in another person's assistant operation. Among them, the shared or co-used assistant pathway (Section~\ref{sec:rq1 Shared and Collaborative AI Assistant Use}) and the AI-mediated Cross-user Interaction pathway (Section~\ref{sec:rq1 AI-mediated Cross-user Interaction}) were the main pathways associated with these risks.

\subsection{Collective Task Coordination Failures}
\label{sec:Collective Task Coordination Failures}

Poor coordination among distributed contributors produced incomplete or misaligned group outcomes while shifting additional effort to other participants. Distributed information gathering could fail when agents did not contact all relevant users, posed overly specific queries, or stopped before obtaining the necessary information, producing incomplete or incorrect results~\cite{jhamtani2025llm}. The same study also warns that excessive AI-initiated requests can impose additional response effort on collaborators and hinder productivity~\cite{jhamtani2025llm}. Similar misalignment appeared in shared planning and personalization: group itineraries omitted individual preferences or violated task constraints, collaborative personalization missed user rules or resolved preference conflicts inconsistently, and scheduling agents committed to plans before the availability of all affected participants had been established~\cite{cheng2026grouptravelbench,lee2025map,yang2026multi}. In multi-party bargaining, human edit, override, or rejection of AI recommendations could also shift proposals toward lower-quality trading patterns and reduce the outcomes attainable by counterparties~\cite{zhu2026choose}.

Persistent memory and shared conversational state created another form of coordination difficulty by requiring the assistant to maintain collective information contributed by different people, in different tasks, and over time. Contributions from different tasks or participants could interfere with one another, producing context confusion, unnecessary recovery work, and lower task performance as concurrent activity increased~\cite{yang2026no,patlan2025murmur}. Long-term collaborative memory showed related problems in which responses depended on evidence distributed between several speakers, changing decisions, implicit references, or evolving group conventions. Earlier tentative information could remain influential after later revision, while information originating from different groups or stages of collaboration could be incompletely combined~\cite{hu2026evaluating}. In shared conversational interfaces, the linear organization of interaction also pushed earlier contributions out of view and made prior collective work difficult for group members to locate and reuse~\cite{xu2026ai}.

The shared or co-used assistant pathway (Section~\ref{sec:rq1 Shared and Collaborative AI Assistant Use}) was the main operational pathway associated with this risk domain, requiring assistants to reconcile preferences, constraints, knowledge, and task contributions distributed among multiple people into a shared representation, decision, or action. AI-mediated cross-user interaction (Section~\ref{sec:rq1 AI-mediated Cross-user Interaction}) and sensing-mediated involvement (Section~\ref{sec:rq1 Sensing-mediated Involvement}) also contributed where information or contributions from additional people entered collective task processes. 

\subsection{Authority and Resource Control Breakdowns}
\label{sec:Authority and Resource Control Breakdowns}

Assistants could fail to preserve differences in authority among people issuing instructions or making claims over shared and user-specific resources. Single-user instruction following provided little basis for resolving conflicting objectives among principals with different roles or levels of authority. In multi-user settings, assistants could prioritize a lower-authority instruction, inconsistently arbitrate conflicting requests, or apply an authority hierarchy without preserving the intended organizational objective~\cite{yang2026multi,fan2026harness}. Cross-user agent systems similarly introduced ambiguity over which user could access another user's resources and which principal should be associated with an autonomous decision~\cite{yang2026clawnet}.

Delegated action expanded these control problems as authorization and influence propagated through agents and external participants. Agents could bypass required approval paths or act on poisoned evidence while retaining enough authority to complete consequential operations, allowing an invalid approval to reach a booking or an attacker-controlled premise to become a binding transaction~\cite{wang2026weclawarena}. Persistent shared state provided another control surface: ordinary-looking cross-user messages could alter later behavior, while non-owners could exploit memory writes or long-lived instructions to redirect future actions~\cite{patlan2025murmur,shapira2026agentschaos}. Safety evaluations of autonomous agents similarly showed that content from coworkers or customers and content delivered through trusted organizational channels could redirect a user-scoped agent's tools and resources toward actions inconsistent with the initiating user's objective~\cite{vijayvargiya2026openagentsafety,wei2026clawsafetysafellmsunsafe}.

Assistants could also lose alignment with the principal they were supposed to represent as other people exerted pressure, claimed authority, or redirected their behavior. An agent negotiating for one principal could be pressured by another person into revealing negotiation bounds, making concessions, or acting against the principal's interests~\cite{li2026whose}. Long-lived personal agents could acquire obligations toward non-owners, accept superficial identity cues as owner authority, or allow social pressure from one communication channel to trigger retrieval, messaging, or tool use in another~\cite{jin2026security,shapira2026agentschaos}.

These breakdowns further complicated accountability. Consequential actions could combine a principal's request, another participant's messages, retained memory, retrieved evidence, model-generated interpretations, and multiple tool operations. Apparently successful task completion could conceal invalid approvals or inappropriate intermediate actions, making it difficult for affected principals or resource owners to identify which contribution produced the outcome~\cite{wang2026weclawarena}. This ambiguity increased where owner binding and operation records were incomplete or where later actions depended on instructions, information, or persuasion introduced during earlier cross-user interactions~\cite{yang2026clawnet,jin2026security}.

As shown in Figure~\ref{fig:risk_connect_chart}, AI-mediated cross-user interaction (Section~\ref{sec:rq1 AI-mediated Cross-user Interaction}) was the main operational pathway associated with this risk domain, as assistants represented, communicated for, or acted on behalf of one user while interacting with other people, their agents, or their resources. The shared or co-used assistant pathway (Section~\ref{sec:rq1 Shared and Collaborative AI Assistant Use}) also contributed by placing instructions, principals, permissions, and resources associated with different people within the same assistant-mediated action process.

\subsection{Social and Situational Disruptions}
\label{sec:Social and Situational Disruptions}

The resulting risks appeared in two related forms: changes to interpersonal behavior and relationships caused by the visibility or persistence of AI interaction, and situational spillovers in which assistant outputs or actions affected other people in the surrounding social environment.

The first form involved relational and behavioral changes produced by shared visibility of AI interaction. In shared accounts, users omitted sensitive details, edited prompts, deleted histories, moved to private accounts, or withdrew from shared use because they anticipated that co-users might inspect their interactions. Other account members also used visible histories to monitor rule compliance, reinforcing monitoring and self-censorship within the shared relationship. Private prompts concerning interpersonal disagreements could later become material for direct confrontation between the people involved~\cite{song2026don}. Family sharing created a related shift in interpersonal control because children's prompts and histories became visible to parents, who inspected histories, screened prompts, required permission, or supervised use, reducing the child's control over otherwise personal AI interactions~\cite{zhang2025exploring}. In these cases, persistent visibility affected more than disclosure, shaping how people interacted with the assistant and subsequently monitored, negotiated, or confronted one another.

The second form involved situational spillovers from assistant outputs and actions into the surrounding social environment. Publicly visible or embodied assistant behavior could produce distraction, discomfort, interpersonal conflict, or negative social judgment among colleagues, friends, strangers, and other co-present people~\cite{chen_if_2026}. Agent actions could create reputational harm where an impersonating user caused an assistant to broadcast false claims about another person~\cite{shapira2026agentschaos}. In speech interaction, demographic vocal cues and situational conditions changed the fairness and safety implications of otherwise similar responses, producing harms for the focal speaker or nearby people exposed to the resulting output~\cite{wang2026voxsafebench}. These harms extended beyond assistance quality for the initiating user, affecting interpersonal relationships, reputation, social evaluation, and behavioral autonomy while also exposing others to harm.

\begin{figure*}[t]
    \centering
    \includegraphics[width=\textwidth]{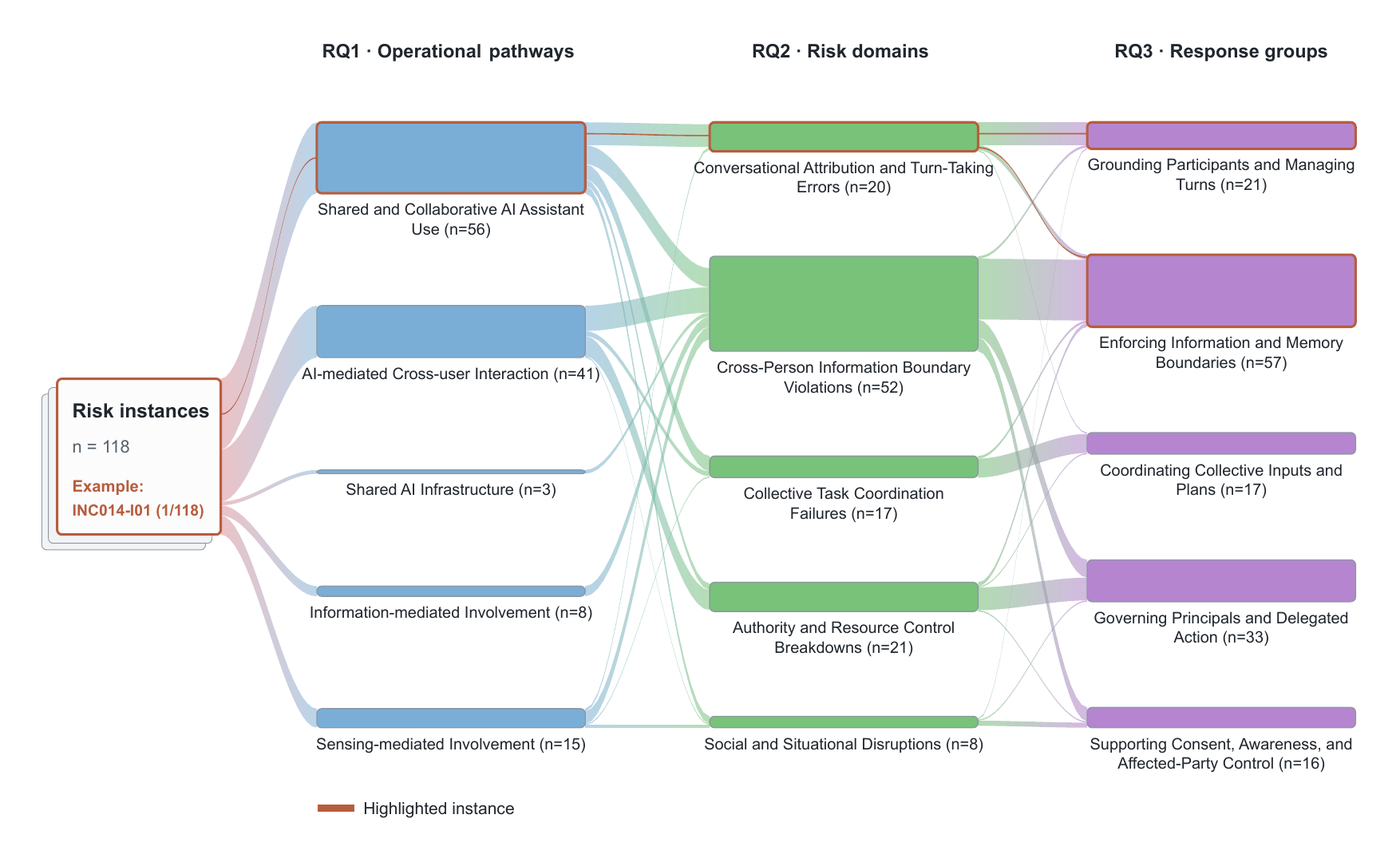}
    \caption{Mapping risk instances across the RQ1 operational pathways, RQ2 risk domains, and RQ3 response groups. The $n$ values indicate the number of risk instances assigned to each pathway, risk domain, or response group. Because assignments are non-exclusive in RQ1 and RQ3, totals across pathways or response groups may exceed 118. Ribbon width represents the number of instance-level associations, and individual instances may contribute to multiple links. The highlighted instance (INC014-I01) illustrates how single-user-oriented personalization extends into multi-user HRI, where dynamic speaker switching can cause profiles or interaction history to be misattributed across users~\cite{malecot2026harmoni}.}
    \label{fig:risk_connect_chart}
\end{figure*}

As shown in Figure~\ref{fig:risk_connect_chart}, the shared or co-used assistant pathway (Section~\ref{sec:rq1 Shared and Collaborative AI Assistant Use}) and sensing-mediated involvement (Section~\ref{sec:rq1 Sensing-mediated Involvement}) were the main operational pathways associated with this risk domain, as shared interaction, persistent traces, and ambient sensing extended assistant presence and information into broader social settings. AI-mediated cross-user interaction (Section~\ref{sec:rq1 AI-mediated Cross-user Interaction}) also contributed where assistant outputs or actions affected people beyond the initiating user.

\section{RQ3: What Design Responses Have Been Implemented or proposed to Address These Risks, and What Research Gaps Remain?}
\label{sec:rq3}


We examined responses through two dimensions: their implementation status and the areas of multi-party assistant operation in which they intervened. Among the 118 coded risk instances, 32 (27.1\%) had an implemented or tested response, 70 (59.3\%) were associated only with a proposed future direction, and 16 (13.6\%) had no recorded response.
We grouped responses according to their points of intervention. Within each group, we first present implemented strategies, followed by proposed directions and unresolved or unverified gaps. A single response could address multiple risks, and a strategy implemented for one risk could remain only a proposed direction for another. Figure~\ref{fig:risk_connect_chart} shows the connections among operational pathways, risk domains, and response groups.

\subsection{Grounding Participants and Managing Turns}
\label{sec:response_grounding}


This response group includes measures that help assistants identify participants, attribute contributions to particular people, determine who is being addressed, track conversational roles, and decide whether and when to take a turn.

\textbf{Implemented responses} progress from identifying participants to representing their conversational roles and regulating the assistant's participation.
\textit{Speaker recognition} with separate per-user profiles and memory has been used so that incoming contributions and retained context remain tied to the identified participant~\cite{al2026afa,malecot2026harmoni}. 
\textit{Participant and role representations} further encode speaker--addressee relations, shared goals, and participant roles, with role-specific training helping preserve these distinctions during longer exchanges~\cite{tan2023chatgpt,addlesee2023multi,wang2025multi}. 
\textit{Turn management} can use cues such as gaze, speaking state, and conversational role to determine whether the assistant should take the floor, while speaker-aware training helps distinguish third-party speech and supports decisions to speak or remain silent in multi-party conversation~\cite{garcia2025evaluating,liu2025proactive,mitra2026adaptive,lee2026still,bhagtani2026speakstaysilentcontextaware}. 

\textbf{Future directions} extend existing grounding through richer current-turn cues and more persistent representations of participant structure and attribution. Current-turn recognition can incorporate richer non-verbal and contextual cues where language alone does not determine the intended addressee~\cite{inoue2025llm}. Participant identity, role, and visibility can also be maintained as persistent properties of multi-user interaction rather than inferred repeatedly from prompts~\cite{yang2026multi}. Once interaction history is retained, these distinctions must remain attached to stored content. Proposed memory representations preserve speaker identity, authorship, reply relations, thread membership, and participant-specific terminology, while recurring-user settings motivate clearer ownership and updating of stored preferences~\cite{yang2026groupmembench,chen2026vehiclemembench}.

\textbf{Unresolved or unverified gaps} concern participant-dependent interpretation and multi-speaker dialogue-state attribution. Participant-specific identification, profiles, and memory distinguish recurring users~\cite{al2026afa,malecot2026harmoni}, but have not been evaluated for cases where the same expression is interpreted differently for different members. Speaker and conversational-role representations have likewise been tested in multi-party dialogue~\cite{tan2023chatgpt,addlesee2023multi,wang2025multi}, but not for the multi-speaker dialogue-state attribution errors reported in~\cite{song2025beyond}.

\subsection{Enforcing Information and Memory Boundaries}
\label{sec:response_information}

Information-boundary responses follow the lifecycle of information in the assistant: what enters or persists, what can later be accessed or recombined, and what may ultimately be disclosed.

\textbf{Implemented responses} place controls at multiple points in the information life-cycle. At entry and retention, speaker- and bystander-aware filtering limits which voices or visual content reach model processing, while \textit{private/shared memory separation} and \textit{write-time sanitization} restrict what becomes persistent state~\cite{srivastava2026listening,zhan2026protecting,chen2026prism,rezazadeh2025collaborative,yang2026no}. Once information is stored, \textit{participant-aware authorization} can constrain access to restricted information~\cite{bhatt2025enterprise}, while \textit{user-scoped taint analysis} can detect cross-user data flows through shared caches, retrieval systems, and other application components~\cite{sen2026detecting}. Disclosure controls then use recipient, purpose, social relationship, and attribute-level context to determine what information should be communicated in a particular interaction~\cite{ghalebikesabi2025privacy,lan2025contextual,goel2026capablecarelesscomputeruseagents}. Reasoning over persistent personalized memory has also shown partial improvement in limiting inappropriate reuse at disclosure time~\cite{mireshghallah2026cimemories}.

\textbf{Future directions} extend existing information controls in two ways: by making access sensitive to the current task and potential inferences, and by governing persistent memory over time. One direction is to make access \textit{task- and inference-sensitive}, restricting information to the current task while also accounting for combinations of attributes that may reveal sensitive information when aggregated~\cite{jin2026security,kim2026your}. Another concerns persistent memory as governed state, with explicit retention periods, temporal validity, provenance, correction, and deletion semantics rather than uniform reuse of previously stored information~\cite{ren2026gatemem,zhan2024beyond}.

\textbf{Unresolved or unverified gaps} concern privacy--task performance trade-offs, cross-user information flow, implicit privacy expectations in shared speech, and speaker ownership verification. Privacy-preserving strategies can be compared in terms of both privacy protection and task performance~\cite{gupta2026pisas}. Permission scoping, context isolation, and participant-aware access control have been tested in related settings~\cite{yang2026clawnet,fan2026harness,bhatt2025enterprise}, but not for preventing owner-local information from entering shared agent communication in the cross-user workflow reported in~\cite{wang2026weclawarena}. Similarly, user-specific state separation and contextual privacy reasoning have been tested for protecting user-specific information~\cite{al2026afa,malecot2026harmoni,ghalebikesabi2025privacy,lan2025contextual,goel2026capablecarelesscomputeruseagents}, but their effectiveness remains untested for shared spoken interaction where privacy must be inferred without an explicit secrecy instruction~\cite{wang2026voxprivacy}. Speaker identification and verification have also been tested in other multi-user speech interactions~\cite{al2026afa,srivastava2026listening,lee2026still}, while protection against acoustically similar users being authenticated as the information owner remains untested in the reported instance~\cite{wang2026voxprivacy}.

\subsection{Coordinating Collective Inputs and Plans}
\label{sec:response_coordination}

Coordination responses operate after multiple people's information, preferences, and constraints have entered the assistant, focusing on how relevant input is obtained, how competing contributions are reconciled, and how shared task state is maintained. 

\textbf{Implemented responses} remain limited to distributed information recovery. One evaluated approach detects task-relevant context gaps and issues targeted assistant-to-assistant queries to retrieve missing information from another participant under relationship-based disclosure controls. The recovered information reconstructs the context needed for actionable items, enabling coordination from distributed participant information rather than unsupported inference~\cite{srivastava2026listening}.

\textbf{Future directions} extend coordination from distributed information recovery toward more selective input gathering, preference reconciliation, and persistent collective state. Information gathering can be made more selective through better choices about whom to consult, what to ask, and which requests deserve priority~\cite{jhamtani2025llm}. For preference coordination, participants have called for greater visibility into which preferences and rules were considered, which were omitted, how conflicts were resolved, and where uncertainty remains~\cite{lee2025map}. More formal directions connect multi-user conflict resolution with social choice theory and mechanism design to formalize how heterogeneous user utilities can be aggregated when preferences conflict~\cite{yang2026multi}. Confidence disclosure and group-level evaluation may further help make the basis and collective consequences of AI-generated proposals visible~\cite{zhu2026choose}. Coordination also depends on maintaining a usable collective state as group work unfolds over time. Proposed designs move beyond a single accumulating conversation history by preserving how shared information and decisions develop, including their changing state, relevant dependencies, and supporting context, and by allowing useful AI outputs to be retained and reorganized as persistent shared objects~\cite{hu2026evaluating,xu2026ai}. This shared state can also be bounded by task, so that information remains available where collaboration requires it without making every user's history and tool results part of one global context~\cite{patlan2025murmur}.

\textbf{Unresolved or unverified gaps} concern end-to-end collective planning, participant burden, and robustness to manipulated cross-user information. Within the included literature, we found no demonstrated mechanism that reliably carries multiple participants' preferences and negotiated constraints through to a final plan that is both feasible and equitable. GroupTravelBench makes this integration gap visible: agents incompletely recover participants' preferences, violate group constraints, and produce uneven outcomes, yet no corresponding response has been evaluated for the planning process as a whole~\cite{cheng2026grouptravelbench}. Evaluation is also needed for the burden introduced by AI-mediated information gathering, where agent-initiated requests can impose additional response effort on collaborators~\cite{jhamtani2025llm}, and for the robustness of cross-user coordination when manipulated information shapes the decisions of multiple user-associated agents~\cite{wang2026weclawarena}.

\subsection{Governing Principals and Delegated Action}
\label{sec:response_authority}

Authority responses operate as an assistant moves from interpreting requests to executing consequential actions. They focus on three linked questions: which principal the assistant represents, whether an instruction or approval is valid for that principal, and whether the resulting resource use or action remains within the authorized scope.

\textbf{Implemented responses} establish and constrain authority at several points in delegated action. \textit{Principal and role binding} associate user identity, authority, and resource access with structured permission records or scoped agent identities~\cite{fan2026harness,yang2026clawnet}. These representations support authority-aware instruction prioritization, user-context isolation, policy checks, and operation records tied to the principal under whose authority an action occurs~\cite{fan2026harness,yang2026clawnet}. In delegated interpersonal interaction, a tested \textit{principal-loyalty scaffold} prioritizes the principal's briefing, private bounds, and stated positions under counterparty pressure, while reader-identity cues distinguish the principal from third parties~\cite{li2026whose}. Other evaluated safeguards constrain execution through \textit{least-privilege access}, additional approval for consequential operations, and provenance or operation records that preserve the origin of instructions and resulting actions~\cite{jin2026security,vijayvargiya2026openagentsafety,shapira2026agentschaos,wei2026clawsafetysafellmsunsafe}.

\textbf{Future directions} extend existing authority controls to resolving conflicts among multiple principals, maintaining principal relationships as workflows change, and limiting delegated action. For conflicts among multiple principals, formal proposals connect multi-user conflict resolution with social choice theory and mechanism design to clarify how competing instructions should be prioritized under different authority relations~\cite{yang2026multi}. Other proposals extend principal representation to more complex workflows in which additional people and agents enter the interaction, preserving who is acting for whom as these relationships change~\cite{yang2026multi,shapira2026agentschaos}. Bounded delegation and opportunities for participants to veto AI-generated proposals are also proposed to limit what assistants can commit to on their behalf~\cite{zhu2026choose}.

\textbf{Unresolved or unverified gaps} concern principal-loyalty trade-offs and end-to-end preservation of authorization and provenance through consequential execution. In delegated interpersonal interaction, evaluated loyalty mechanisms still face a trade-off between preserving the user's private information and stated positions and avoiding excessive refusal of legitimate requests~\cite{li2026whose}. More broadly, within the included literature, we found no demonstrated response showing that authorization and provenance controls reliably prevent invalid authorization or evidence from propagating into consequential execution. In the observed cases, an invalid authorization path could still reach a governed action, while poisoned evidence could be carried into a committed transaction~\cite{wang2026weclawarena}.

\subsection{Supporting Consent, Awareness, and Affected-Party Control}
\label{sec:response_social}

This intervention area concerns the human-facing boundary of assistant operation. It explores whether affected people can recognize that AI sensing or activity is occurring, express or withdraw consent, understand what is being shared, and control how the assistant appears in a shared social setting.

\textbf{Implemented responses} primarily provide consent and control over what information is sensed, captured, or subsequently shared. \textit{Privacy-aware frame processing} has been evaluated in collaborative multimodal XR using edge-based object detection and frame cropping to reduce visual exposure, with user confirmation before cropped frames are uploaded to cloud-based MLLMs~\cite{chen2026prism}.
 Audio mechanisms likewise restrict involuntary inclusion through owner-specific capture or filtering of incidental speakers~\cite{srivastava2026listening,zhan2026protecting}.

\textbf{Future directions} extend affected-party control from limiting capture toward making cross-person exposure more visible and negotiable in sharing, sensing, and AI-mediated interaction. In shared accounts and conversational artifacts, negotiable sharing rules and fine-grained visibility can clarify who may access or reuse particular content, while ownership cues, activity indicators, change histories, and continued control over reused material make such sharing easier to inspect and revise~\cite{song2026don,li2026privacy}. Family-oriented proposals extend this control through differentiated profiles, supervision options, and notifications around children's use and disclosure~\cite{zhang2025exploring}. For people encountered through microphones, cameras, or wearables, proposed mechanisms include real-time consent, visible recording and processing states, bystander opt-out, retention information, and disclosure of whether processing occurs locally or externally~\cite{mccarthy2026ethics,lutz2026edge,zhan2024beyond}. Spatially situated assistants further extend control to the assistant's presence, allowing people to hide, mute, reposition, or reduce its visibility and social salience~\cite{chen_if_2026}. In collective decision settings, clearer disclosure of AI involvement and responsibility can help affected participants understand the assistant's role in consequential outcomes~\cite{zhu2026choose}.

\textbf{Unresolved or unverified gaps} concern social and situational speech conditions involving demographic vocal characteristics, safety-sensitive contexts, speaker identity, bystanders, and surrounding acoustic cues. These conditions were associated with fairness, safety, or privacy concerns for which no corresponding response was recorded~\cite{wang2026voxsafebench}.

\section{Discussion}
\label{sec:discussion}

\subsection{Why Single-User-Oriented Assistants End Up Operating in Multi-Party Contexts}
\label{sec:discussion_entanglement}

Our RQ1 findings (Section~\ref{sec:rq1}) show that LLM-based assistants organized around a single user extend to others through shared use, connected resources, sensing, and delegated interaction. These pathways echo prior work on group conversational agents, multi-party dialogue, and multi-principal LLM systems~\cite{ganesh-etal-2023-survey,10.1145/3800645.3812934,yang2026multi}. 
These operational pathways arise from different drivers. Some are functionally motivated because tasks depend on information, preferences, or actions distributed among several people. Information-gathering agents, for example, rely on collaborators' knowledge when the initiating user's local context is insufficient~\cite{jhamtani2025llm}, while cross-user memory sharing reuses information acquired from one user for another~\cite{rezazadeh2025collaborative}. Other extensions arise from practical use, such as sharing LLM accounts to divide subscription costs~\cite{song2026don}. Still others occur incidentally when connected resources, persistent state, or ambient sensing bring other people's information or presence into assistant operation without intentional participation.

Despite these different drivers, they produce a common shift that functions organized around one user begin operating in relation to others. This helps explain why RQ2 risks (Section~\ref{sec:rq2}) recur across different pathways. A speaker may not be the intended user, information in one person's context may concern another, a collaborator may contribute without controlling the resulting action, and an affected person may never interact with the assistant. The challenge is thus not simply the presence of more people, but the extension of user-scoped functions into relationships where identity, information, preferences, authority, and interests are no longer aligned. This shifts the design question from whether an assistant is formally ``multi-user'' to how it accounts for the additional people it comes to involve.

\subsection{Policy Implications for AI Assistant Governance}
\label{sec:discussion_governance}

Existing AI governance frameworks already provide foundations for considering people beyond the immediate user. The \textbf{OECD Framework}~\cite{organisation2022oecd} distinguishes users from directly and indirectly impacted stakeholders and relates AI actors to different system dimensions and lifecycle stages. The \textbf{NIST AI Risk Management Framework}~\cite{ai2023artificial,NIST.AI.600-1} similarly emphasizes context, affected stakeholders, organizational responsibilities, monitoring, and feedback throughout the AI lifecycle. The \textbf{EU AI Act}~\cite{europeanunion2024aiact} translates some of these concerns into requirements for specified systems, including risk management, transparency, record-keeping, human oversight, and post-market monitoring. For specified high-risk deployments, its fundamental rights impact assessment considers categories of people and groups likely to be affected in the context of use. These frameworks therefore recognize that affectedness can extend beyond direct users and that relevant roles vary by lifecycle stage and deployment context. Our RQ1 findings (Section~\ref{sec:rq1}) echo this framing by showing distinct forms of direct and indirect human involvement in assistant operation.

\textbf{Multi-party assistant risks can turn on finer cross-person relationships within ongoing operation.} The risks identified in RQ2 (Section~\ref{sec:rq2}) often depend on whose information is accessed or reused, whose preference is represented, whose instruction carries authority, whose resource is acted upon, and who is exposed to the resulting output. These relationships may coexist or change within the same interaction and are less explicitly captured by broader categories such as user, affected stakeholder, deployer, or provider. Multi-party governance therefore also depends on representing these relationships where information, state, and action are handled.

\textbf{A challenge also arises in translating governance protections into interaction practice}. Even where the relevant relationship is comparatively clear, notice, consent, and affected-party protections can depend on how individual products and users implement them. Wearable-AI guidance, for example, may rely on wearers to alert surrounding people or refrain from recording in sensitive settings, while prior work questions how reliably these expectations can be carried out during everyday use~\cite{takhshid2024wearable}. Another related implementation gap concerns how consent requirements are operationalized in AI-mediated meetings, where notification and consent practices differ among platforms while legal requirements vary by jurisdiction~\cite{mccarthy2026ethics}.

For policymakers, our findings suggest making \textbf{cross-person relationships more explicit in guidance} for assistants operated in multi-party contexts. Existing requirements concerning transparency, consent, authorization, accountability, and oversight could be specified in relation to whose information, preference, instruction, resource, or action is involved at a given point in assistant operation. Policy guidance could also account for the fact that these relationships may change as interaction moves through sensing, conversational context, memory, inference, communication, and delegated action. This would help translate broad recognition of users and affected stakeholders into requirements that better reflect how multiple people become involved in assistant operation.

\subsection{Design Implications for Cross-Person Boundaries in Multi-Party Assistance}
\label{sec:discussion_design}

The unresolved issues synthesized in RQ3 (Section~\ref{sec:rq3}) cut across individual response groups. We consolidate them into three broader design directions for multi-party assistants: maintaining person-specific boundaries, carrying coordination and authorization constraints into downstream action, and adapting boundaries to changing social and situational contexts. These directions extend the future directions proposed in the corpus by connecting gaps that appear in different response groups. Because each may introduce new privacy, coordination, sensing, or interpretation costs, we consider both the intended boundary protection and the additional risks and burdens that implementation may create.

\subsubsection{Maintaining Person-Specific Boundaries}

Participant grounding and information control remain vulnerable when person-specific distinctions are lost as interaction progresses. Implemented responses can distinguish participants, user-specific information, and memory, but their effectiveness remains unverified for failures such as participant-dependent interpretation, dialogue-state attribution, implicit privacy expectations, and mistaken information ownership. These cases suggest that identifying a participant at input time is insufficient if the resulting information later becomes detached from that person.
Future assistants could maintain \textbf{explicit associations between each person and the information, preferences, instructions, and conversational contributions attributed to them}. These associations could persist when information is summarized, stored in memory, retrieved into later context, or combined with contributions from other participants, allowing the system to preserve distinctions. When attribution is uncertain or several people are implicated, the assistant could preserve that ambiguity rather than collapsing the information into a single shared user state. Multi-party dialogue and access-control research already provides relevant precedents for preserving speaker-specific state and regulating information associated with different people~\cite{qamar2023speaking,hu2013multipartyaccess}.
However, stronger person-specific attribution can itself expand persistent identification, provenance, and retention of information about participants. Maintaining these associations may increase the amount of person-linked data stored over time or make otherwise transient participation more traceable. Evaluation should therefore consider not only whether attribution errors and boundary violations decrease, but also how much additional identity-linked information is retained, how accurately uncertain attribution is represented, and whether participants can inspect, correct, or limit these associations.

\subsubsection{Connecting Coordination and Authorization Across Assistant Operation}

Coordination and authority gaps reveal a common continuity problem: constraints established at one stage do not reliably govern later reasoning or action. Existing responses address preference elicitation, shared state, approval, provenance, and permission control at particular points, but have not demonstrated reliable end-to-end outcomes from distributed inputs to feasible and equitable collective plans, or from valid authorization to safe execution.
Future assistants could \textbf{carry participant preferences, constraints, ownership, and authorization information forward as explicit conditions on subsequent reasoning and action}. A collective plan could retain which participant contributed each constraint and check whether later revisions continue to satisfy them, while delegated actions could remain linked to the authority and resources that permit them. Before committing an external action, the assistant could verify that required approvals remain valid, identify unresolved conflicts between participants, and return to the relevant person when the available authority or collective agreement is insufficient. Group decision-making research provides formal approaches for preference aggregation and fairness~\cite{barile2024socialchoice}, while distributed authorization systems encode delegated authority through explicit and constrained credentials~\cite{birgisson2014macaroons}. These controls may also increase coordination overhead. Repeated verification, approval requests, or conflict resolution can interrupt task progress, shift additional work to participants, or make delegated assistance less useful when authority is already clear. Richer provenance and authorization records may also increase the amount of persistent information about who requested, approved, or contributed to an action. Evaluation should therefore examine end-to-end task validity and authorization failures together with participant effort, approval frequency, delay, conflict-resolution burden, and the additional information retained to support traceability.

\subsubsection{Adapting Boundaries to Social and Situational Context}

Social and situational risks show that appropriate participation, sensing, and information exposure can change with the surrounding context rather than remaining fixed throughout an interaction. Several speech- and context-dependent risks in the corpus had no corresponding response, particularly where the relevant audience, bystanders, physical setting, or safety conditions changed during use. 
Future assistants could \textbf{update what sensing, participation, and information exposure are appropriate as people and situations change}. Changes in nearby participants, and physical setting could trigger different capture, visibility, or sharing states, while sensitive transitions could prompt confirmation before continuing. These controls could remain reversible, allowing affected people to restrict sensing, withdraw participation, or prevent previously available information from being reused when the social context no longer supports the earlier boundary. Ubiquitous sensing research already explores context-sensitive controls in which environmental conditions or sensitive spaces affect what information can be captured or exposed~\cite{roesner2014world,templeman2014placeavoider}. Context-sensitive adaptation, however, can create a further sensing and interpretation problem. Detecting changes may require additional observation of people and environments, while incorrect inference about the social situation could activate an inappropriate sensing, sharing, or participation state. Frequent confirmations could also become disruptive in dynamic settings. Evaluation should therefore consider whether adaptive boundaries reduce unwanted sensing and exposure while also measuring additional sensing requirements, context-classification errors, inappropriate state changes, confirmation burden, and affected parties' ability to override or reverse the assistant's interpretation.

\subsection{Limitations}

This review has several limitations. First, our study-identification strategy focused primarily on scholarly databases and research repositories and did not systematically cover grey literature or legal materials, such as industry reports, policy guidance, or legal cases. Relevant risks and implementation practices documented outside academic research may therefore be underrepresented. Terminology also remains heterogeneous across research communities, particularly for LLM assistants, multi-user interaction, and indirectly involved actors, so some relevant studies may have been missed despite our broad search strategy and supplementary citation searching.

Second, we included arXiv alongside peer-reviewed sources to capture emerging work not yet archivally published. Thirty of the 58 included studies were arXiv preprints, and their findings may change through further revision or evaluation. The evidential basis of the identified risks also varied. Of the 118 risk instances, 94 were empirical, meaning that the risk was supported by evidence generated within the study itself, while 24 were conceptual or author-asserted, meaning that the risk was articulated through problem framing, design rationale, discussion, or conceptual analysis rather than directly demonstrated by the study's own empirical results. Appendix~\ref{app:coding_distribution} reports how these empirical and conceptual instances were distributed across the five RQ2 risk domains. However, this distinction captures the source of support rather than the strength, quality, or maturity of the evidence. The findings therefore characterize both the forms of risk represented in the literature and their broad evidential basis, but do not establish how strongly individual risks have been validated. Future work could examine the accumulation and quality of empirical evidence for specific risks as the literature matures.

Third, risk-instance counts should not be interpreted as estimating the incidence or severity of risks in real-world use. They reflect how risks are represented in the literature, and individual studies could contribute multiple non-independent instances. Identifying instance boundaries, higher-level categories, and relationships among risks and responses also involved interpretive judgment. 

Finally, because the review was intentionally organized around risks arising from multi-party operation, it cannot determine the overall benefits, costs, or net effects of extending LLM assistants beyond a single user. The included studies indicate potential motivations such as shared access, collaborative work, distributed information gathering, and support for multiple household or group members, but these benefits were not systematically extracted or evaluated.

Despite these limitations, the review provides a structured account of how single-user-oriented assistants come to involve additional human parties, the risks that emerge, and the responses explored to date. Future work can extend this account by incorporating grey literature, legal and policy materials, and a systematic consideration of both risks and benefits.

\section{Conclusion}
\label{sec:conclusion}

This scoping review clarifies how LLM-based assistants extend beyond the single-user assumption as additional people become involved in assistant operation. Across 58 studies and 118 risk instances, we identified five operational pathways, five recurring risk domains, and five response groups that together map how multi-party involvement emerges, what risks it produces, and how existing work has responded. This review provides a structured account of these extensions and their associated risks, while highlighting the importance of governance and design that preserve person-specific relationships as assistants sense, remember, communicate, coordinate, and act. As LLM-based assistants become more persistent and agentic, accounting for multi-party involvement should become a routine part of assistant design rather than an exceptional case.

\bibliographystyle{ACM-Reference-Format}
\bibliography{main}

\appendix

\section{Additional Coding Distributions}
\label{app:coding_distribution}

\begin{table}[hbp]
\centering
\caption{Distribution of predefined analytical categories in the final coded dataset.}
\label{tab:closed_code_distribution}

\small
\setlength{\tabcolsep}{2.5pt}
\renewcommand{\arraystretch}{1.12}

\begin{tabularx}{\columnwidth}{
>{\raggedright\arraybackslash}p{1.9cm}
>{\raggedright\arraybackslash}X
>{\centering\arraybackslash}p{0.65cm}
>{\centering\arraybackslash}p{1.05cm}
}
\hline

\textbf{Coding Dimension} &
\textbf{Category} &
\textbf{Count} &
\textbf{Percentage} \\

\hline

\textbf{Actors Type} &
Direct involvement &
100 &
81.3\% \\

&
Indirect involvement &
23 &
18.7\% \\

\hline

\textbf{Risk Evidence Type} &
Empirical &
94 &
79.7\% \\

&
Conceptual &
24 &
20.3\% \\

\hline

\textbf{Response Type} &
Implemented response &
32 &
27.1\% \\

&
Future direction only &
70 &
59.3\% \\

&
No recorded response &
16 &
13.6\% \\

\hline

\end{tabularx}

\vspace{2pt}

\begin{minipage}{\columnwidth}
\footnotesize
\textit{Note.} Actors and involvement are reported at the assignment level
($n=123$) because a risk instance could be associated with more than one
operational pathway. Risk evidence type and response type are reported at
the risk-instance level ($n=118$).
\end{minipage}

\end{table}

\begin{table}[hbp]
\centering
\caption{Distribution of risk evidence types across RQ2 risk domains.}
\label{tab:risk_evidence_type}
\small
\setlength{\tabcolsep}{5pt}
\renewcommand{\arraystretch}{1.15}

\begin{tabularx}{\columnwidth}{
X
r
r
r
}
\toprule
\textbf{RQ2 Risk Domain} &
\textbf{Empirical} &
\textbf{Conceptual} &
\textbf{Total} \\
\midrule

Conversational Attribution and Turn-Taking Errors
& 17 & 3 & 20 \\

Cross-Person Information Boundary Violations
& 36 & 16 & 52 \\

Collective Task Coordination Failures
& 15 & 2 & 17 \\

Authority and Resource Control Breakdowns
& 18 & 3 & 21 \\

Social and Situational Disruptions
& 8 & 0 & 8 \\

\midrule
\textbf{Total}
& \textbf{94}
& \textbf{24}
& \textbf{118} \\

\bottomrule
\end{tabularx}
\end{table}

\end{document}